\documentclass[stef]{stefano}

\usepackage{amsmath}
\usepackage{amssymb}
\usepackage{graphics}
\usepackage{rotating}
\usepackage{cite}
\usepackage{color}
\usepackage{fancybox}

\def\makeheadbox{{
\begin{flushright}{\fcolorbox{black}{yellow}
{\color{red} $\bullet$ {\color{black}{ {\sc Journal of
Mathematical Physics} {\bf 51}, 113504-10 (2010) }}
{\color{red}{$\bullet$}} }}
\end{flushright}
}}

\def\r{\boldsymbol{r}}
\def\0{\mbox{\tiny $0$}}
\def\1{\mbox{\tiny $1$}}
\def\2{\mbox{\tiny $2$}}
\def\3{\mbox{\tiny $3$}}
\def\4{\mbox{\tiny $4$}}
\def\5{\mbox{\tiny $5$}}
\def\6{\mbox{\tiny $6$}}
\def\7{\mbox{\tiny $7$}}
\def\8{\mbox{\tiny $8$}}
\def\9{\mbox{\tiny $9$}}
\def\I{\mbox{\tiny I}}
\def\II{\mbox{\tiny II}}
\def\III{\mbox{\tiny III}}
\def\mi{\mbox{\tiny $-$}}
\def\pl{\mbox{\tiny $+$}}

\def\pem{\mbox{\tiny $\pm$}}

\begin{document}
%

\title{\large \shadowbox{
\begin{tabular}{c} A CLOSED FORMULA FOR THE BARRIER TRANSMISSION COEFFICIENT\\
IN QUATERNIONIC QUANTUM MECHANICS\end{tabular}}}

\author{Stefano De Leo\inst{1}
\and Gisele Ducati\inst{2}
\and Vinicius Leonardi\inst{1} \and Kenia Pereira\inst{1}. }

\institute{
Department of Applied Mathematics, State University of Campinas,\\
SP 13083-970, Campinas, Brazil,\\
{\em deleo@ime.unicamp.br}.
\and
CMCC, Federal University of ABC,\\
SP 09210-170, Santo Andr\'e, Brazil,\\
{\em ducati@ufabc.edu.br}.
}


\date{Submitted: {\em May, 2010}.  Revised: {\em Sepetember, 2010}. }

\abstract{In this paper, we analyze, by using a matrix approach, the dynamics
of a non-relativistic particle in presence of a quaternionic potential
barrier. The matrix method used to solve the quaternionic Schrödinger equation
allows to obtain a closed formula for the transmission coefficient. Up to now,
in quaternionic quantum mechanics, almost every discussion on the dynamics of
non-relativistic particle was motived by or evolved from numerical studies. A
closed formula for the transmission coefficient stimulates an analysis of {\em
qualitative} differences between complex and quaternionic quantum mechanics,
and, by using the stationary phase method, gives the possibility to discuss
transmission times.}










\titlerunning{\sc transmission coefficient for quaternionic barriers}

\maketitle


\section*{I. INTRODUCTION}
In the quaternionic formulation of non-relativistic quantum
mechanics, the dynamics of a particle without spin subject to the
influence of the anti-hermitian scalar potential,
\[ i\,V_{\1}(\r, t) + j\,V_{\2}(\r, t) +k\,V_{\3}(\r, t)\,\,, \]
is described by
\begin{equation}
\label{q1} \hbar\,\partial_{t} \Psi( \r,t) = \left[ \,
i\,\frac{\,\,\hbar^{^{2}}}{2m} \, \nabla^{^{2}} - i\, V_{\1}(\r,
t) - j \, V_{\2}(\r, t) - k \, V_{\3}(\r, t) \, \right] \,
\Psi(\r,t)~.
\end{equation}
Eq.(\ref{q1}) is known as the quaternionic  Schr\"odinger equation. For a
detailed discussion on foundations of quaternionic quantum mechanics, we refer
the reader to Adler's book\cite{ADL}. Numerical investigations
 on the observability
of quaternionic deviations from the standard (complex) quantum
mechanics\cite{PER79,KAI84,KLE88,DAV89,DAV92,DDN02,DD05,DD06,DD07} have
recently stimulated the study of new mathematical tools in solving
quaternionic differential
equations\cite{DeSco00,DeDuc01,DeScoSol02,DeDuc03,DeDuc04,DeDuc06}.

In complex quantum mechanics\cite{COHEN,MER}, the rapid spatial
variations of a square potential introduce purely quantum effects
in the motion of the particle. Before beginning our analytic study
of the quaternionic potential barrier and analyze analogies and
differences between the complex and quaternionic dynamics, we
introduce some important mathematical properties of the
quaternionic Schr\"odinger equation in the presence of {\em time
independent} potentials,
\begin{equation}
\label{q2} \hbar\,\partial_{t} \Psi( \r,t) = \left[ \,
i\,\frac{\,\,\hbar^{^{2}}}{2m} \, \nabla^{^{2}} - i\, V_{\1}(\r) -
j \, V_{\2}(\r) - k \, V_{\3}(\r) \, \right] \, \Psi(\r,t)~.
\end{equation}
Due to the fact that our analysis is done  for a time independent potential,
we can apply the method of separation of variables. So, we introduce
\begin{equation}
\label{ss} \Psi(\r, t) = \Phi(\r) \, \exp[ \, - \, \mbox{$\frac{i}{\hbar}$} \,
E \, t \, ]\,\,.
\end{equation}
The position on the right of the complex exponential is important to factorize
the time dependent function. Then, Eq.(\ref{q2}) reduces to the following
quaternionic (right complex linear)  ordinary differential
equation\cite{DeDuc01}
\begin{equation} \label{q3}
-\,E\,\Phi(\r)\,i = \left[ \, i\,\frac{\,\,\hbar^{^{2}}}{2m} \, \nabla^{^{2}}
- i\, V_{\1}(\r) - j \, V_{\2}(\r) - k \, V_{\3}(\r) \, \right] \, \Phi(\r)~.
\end{equation}
In the case of a one-dimensional potential,
$\boldsymbol{V}(\r)=\boldsymbol{V}(x)$, and for a particle moving in the
$x$-direction,  $\Phi(\r)=\Phi(x)$, the previous equation becomes
\begin{equation} \label{q3}
-\,E\,\Phi(x)\,i = \left[ \, i\,\frac{\,\,\hbar^{^{2}}}{2m}\,
\partial_{xx} - i\, V_{\1}(x) - j \, V_{\2}(x) - k \, V_{\3}(x) \, \right]
\, \Phi(x)~.
\end{equation}
In the case of a quaternionic barrier,
\[ \boldsymbol{V}(x)=\left\{\,\mbox{zone I}:\,\,\boldsymbol{0}\,\,\,(x<0)\,\,,
\,\,\,\,\,\mbox{zone II}:\,\,\boldsymbol{V}\,\,\,(0<x<L)\,\,,
\,\,\,\,\,\mbox{zone
III}:\,\,\boldsymbol{0}\,\,\,(x>L)\,\right\}\,\,,\]
 the plane wave solution is
obtained by solving a second order differential equations with (left) constant
quaternionic coefficients\cite{DeDuc01,DeDuc06}. To shorten our notation, it
is convenient to rewrite the quaternionic differential equation (\ref{q3}) in
terms of adimensional quantities. In the potential region, $0<\xi<\lambda$,
the differential equation to be solved is
\begin{equation} \label{qad}
i\,\Phi_{\II}''(\xi)= i\, V_{c}\,\Phi_{\II}(\xi) + j \,
V_{q}\,e^{-i\theta}\,\Phi_{\II}(\xi)-\epsilon^{\2}\,\Phi_{\II}(\xi)\,i\,\,,
\end{equation}
where
\[\epsilon=\sqrt{E/V_{\0}}\,\,,\,\,\,
V_{\0}=\sqrt{V_{\1}^{^{\2}}+V_{\2}^{^{2}}+V_{\3}^{^{2}}}\,\,,\,\,\,
V_c=V_{\1}/V_{\0}\,\,,\,\,\,
V_q=\sqrt{V_{\2}^{^{2}}+V_{\3}^{^{2}}}/V_{\0}\,\,,\,\,\, \arctan
\theta = V_{\3}/V_{\2}\,\,,
\]
and
\[
\xi = \sqrt{2m\,V_{\0}/\hbar^{^{2}}}\,\,x\,\,,\,\,\,\,\,
\lambda = \sqrt{2m\,V_{\0}/\hbar^{^{2}}}\,\,L\,\,.
\]
 The general solution contains
 four complex coefficients ($A$, $\widetilde{A}$, $B$, and  $\widetilde{B}$) to be
 determined by the
boundary conditions,
\begin{equation}
\label{solg} \Phi_{\II}(\xi) =   ( 1 + j\, \gamma) \,  \left\{
\exp \left[ \, \alpha_{\mi} \, \xi \, \right] \, A + \exp \left[
\, - \, \alpha_{\mi} \, \xi \, \right] \, B \right\} +  ( \beta +
j ) \, \left\{ \exp \left[ \, \alpha_{\pl} \, \xi \, \right] \,
\widetilde{A} + \exp \left[ \, - \, \alpha_{\pl} \, \xi \, \right]
\, \widetilde{B} \right\}\,\,,
\end{equation}
for a detailed derivation of this solution see ref.\cite{DeDuc06}. The
potential and energy inputs appear in the following complex quantities
\begin{equation}
\alpha_{\pem}  =  \sqrt{V_{c} \pm \sqrt{\epsilon^{\4} -
V_{q}^{^{2}}}} \,\,\,,\,\,\,\,\, \beta=i\,\frac{V_q\,e^{i\theta}}{
\epsilon^{\2} + \sqrt{\epsilon^{\4} - V_{q}^{^{2}}}}\,\,\,,
\,\,\,\,\,\gamma=-\,i\,\frac{V_q\,e^{-i\theta}}{\epsilon^{\2} +
\sqrt{\epsilon^{\4} - V_{q}^{^{2}}}}\,\,\,.
\end{equation}
Note that for $E\geq\sqrt{V_{\2}^{^{2}}+V_{\3}^{^{2}}}$, we have
$\gamma=\beta^*$.

 In the free potential regions, zone I ($\xi<0$) and zone III ($\xi>\lambda$), the
solutions are given by
\begin{equation}
\label{free}
\begin{array}{lcl}
\Phi_{\I}(\xi) & = & \exp \left[ \, i \, \epsilon \, \xi \, \right]  + \exp
\left[ \, - \, i \,  \epsilon  \, \xi \, \right] \, R + j \, \exp \left[ \,
 \epsilon  \, \xi \, \right] \, \widetilde{R}\,\,, \\
\Phi_{\III}(\xi) & = & \exp \left[ \, i \, \epsilon \, \xi \,
\right] \, T + j \,
 \exp \left[ \, - \,
 \epsilon  \, \xi \,  \right] \, \widetilde{T}\,\,.
 \end{array}
\end{equation}
It is important to observe that starting from theses solutions we can  find
the relation between reflection and transmission coefficients without the
necessity to know their explicit formulas. In fact,  for stationary states the
density probability is independent from the time and, consequently,  the
current density $\Phi^{\dag}(x)\,i\,\Phi'(x)+\mbox{h.c.}$ is a constant in
$x$. This implies, for example,
\[
\Phi_{\I}^{\dag}(0)\,i\,\Phi_{\I}'(0)+\mbox{h.c.} =
\Phi_{\III}^{\dag}(\lambda)\,i\,\Phi_{\III}'(\lambda)+\mbox{h.c.}\,\,.
\]
From this relation, we get the well known  norm conservation formula, i.e.
\begin{equation}
\label{nc} 1 - |R|^{^{2}} = |T|^{^{\2}}\,\,.
\end{equation}
Having introduced this preliminary discussion on the quaternionic
Schr\"odinger equation, we can now return to our task of finding a closed
formula for the transmission coefficient. Once obtained the transmission
coefficient it suffices to use Eq.(\ref{nc}) to find the norm of the reflected
wave.

\section*{II. CONTINUITY CONSTRAINTS AND MATRIX EQUATION}
To find a closed formula for the transmission coefficient, we impose the
continuity of $\Phi(\xi)$ and its derivative, $\Phi'(\xi)$, at the points
where the potential is discontinuous, i.e.
 \[
 \Phi_{\I}(0)=\Phi_{\II}(0)
 \,\,,\,\,\,\,\,
\Phi_{\I}'(0)=\Phi_{\II}'(0) \,\,,\,\,\,\,\,
\Phi_{\III}(\lambda)=\Phi_{\II}(\lambda)
 \,\,,\,\,\,\,\,
\Phi_{\III}'(\lambda)=\Phi_{\II}'(\lambda)\,\,.
\]
Using these constraints and  separating the complex from the pure
quaternionic part, we find
\begin{eqnarray}
\label{tmat}
1+R&=&A+B+\beta\,(\widetilde{A}+\widetilde{B})\,\,,\nonumber
\\
i\,\frac{\epsilon}{\alpha_{\mi}}\,\,(1-R)&=&A-B
+\,\frac{\alpha_{\pl}}{\alpha_{\mi}}\,\,\beta\,
(\widetilde{A}-\widetilde{B})\,\,,\nonumber\\
\widetilde{R}&=&\gamma\,(A+B)+
\widetilde{A}+\widetilde{B}\,\,,\nonumber
\\
\frac{\epsilon}{\alpha_{\mi}}\,\,\widetilde{R}&=&\gamma\,(A-B)+
\frac{\alpha_{\pl}}{\alpha_{\mi}}\,\,
(\widetilde{A}-\widetilde{B})\,\,,\nonumber\\
T\,e^{i\epsilon \lambda} & = & A\,e^{\alpha_{\mi} \lambda} +
B\,e^{-\,\alpha_{\mi} \lambda} +
\beta\,(\widetilde{A}\,e^{\alpha_{\pl} \lambda} +
\widetilde{B}\,e^{-\,\alpha_{\pl} \lambda})\,\,,\nonumber \\
i\,\frac{\epsilon}{\alpha_{\mi}}\,\,T\,e^{i\epsilon \lambda} & = &
A\,e^{\alpha_{\mi} \lambda} - B\,e^{-\,\alpha_{\mi} \lambda} +
\,\frac{\alpha_{\pl}}{\alpha_{\mi}}\,\,
\beta\,(\widetilde{A}\,e^{\alpha_{\pl} \lambda} -
\widetilde{B}\,e^{-\,\alpha_{\pl} \lambda})\,\,,\nonumber \\
\widetilde{T}\,e^{-\,\epsilon \lambda} & = &
\gamma\,(A\,e^{\alpha_{\mi} \lambda} + B\,e^{-\,\alpha_{\mi}
\lambda}) + \widetilde{A}\,e^{\alpha_{\pl} \lambda} +
\widetilde{B}\,e^{-\,\alpha_{\pl} \lambda}\,\,,\nonumber \\
-\,\frac{\epsilon}{\alpha_{\mi}}\,\,\widetilde{T}\,e^{-\,\epsilon
\lambda} & = & \gamma\,( A\,e^{\alpha_{\mi} \lambda} -
B\,e^{-\,\alpha_{\mi} \lambda})
+\,\frac{\alpha_{\pl}}{\alpha_{\mi}}\,\,
(\widetilde{A}\,e^{\alpha_{\pl} \lambda} -
\widetilde{B}\,e^{-\,\alpha_{\pl} \lambda})\,\,.
\end{eqnarray}
The procedure for calculating reflection and transmission
coefficients by using the transfer matrix method is a  well-known
technique both in quantum mechanics\cite{COHEN,MER} and
optics\cite{HAV,BORN}. It is based on the fact that, according to
Schr\"odinger or Maxwell equations, there are simple continuity
conditions for the field across boundaries from one medium to the
next. If the field is known at the beginning of a layer, the field
at the end of the layer can be derived from a simple matrix
operation. The final step of the method involves converting the
system matrix back into reflection and transmission coefficients.
From Eq.(\ref{tmat}),  after straightforward algebraic
manipulations, we find
\begin{equation}
\label{transreflec} \left(\begin{array}{c} 1 \\ R \\ \widetilde{R} \\
\widetilde{R}
\end{array} \right) =\,
F^{^{-1}}\,\,\underbrace{\,G\,\Delta_{\lambda}\,G^{^{-1}}}_{\mbox{\large
$M$}}\,\,
\left(\begin{array}{rl} T &e^{i\,\epsilon \lambda} \\
i\,\frac{\epsilon}{\alpha_{\mi}}\,\,T & e^{i\,\epsilon \lambda} \\
\widetilde{T} & e^{-\,\epsilon \lambda}
\\-\,\frac{\epsilon}{\alpha_{\mi}}\,\,\widetilde{T} & e^{-\,\epsilon \lambda}
\end{array} \right)\,\,,
\end{equation}
where
\[ F= \left(\begin{array}{cccc} 1 & 1 & \,\,0\,\, & 0 \\
i\,\frac{\epsilon}{\alpha_{\mi}} & \,\,\,\,- \,
i\,\frac{\epsilon}{\alpha_{\mi}}\,\,\,\, & 0 & 0
\\ 0 & 0 & 1 & 0 \\ 0 & 0 & 0 & \,\,\frac{\epsilon}{\alpha_{\mi}}\,\, \end{array}
\right)\,\,\,,\,\,\,\,\,\,\,
G=\left(\begin{array}{rrrr}  1 & 1 & \beta &  \beta \\
1 & \,\,\,\,-1 & \,\,\,\,\beta\,\frac{\alpha_{\pl}}{\alpha_{\mi}}
&
\,\,\,\,-\,\beta\,\frac{\alpha_{\pl}}{\alpha_{\mi}}\\
\gamma & \gamma & 1 & 1 \\
\gamma & -\gamma & \frac{\alpha_{\pl}}{\alpha_{\mi}} & -\,
\frac{\alpha_{\pl}}{\alpha_{\mi}}
\end{array} \right)\,\,,\]
and
\[
\Delta_{\lambda}=\mbox{diag}\{\, e^{-\,\alpha_-\lambda}\,,\,
e^{\alpha_-\lambda}\,,\, e^{-\,\alpha_+\lambda}\,,\,
e^{\alpha_+\lambda}\,\}\,\,.
\]
From this matrix system, we can extract the transmission coefficient $T$. The
closed formula is not simple and we prefer to give it  in terms of the matrix
elements of $M$,
\begin{equation}
\label{tram}
 T =
2\,\exp[-\,i\,\epsilon\,\lambda]\,/\,\mathcal{D}\,\,,
\end{equation}
where $\mathcal{D}$ is given by
\[
M_{\1 \1}+\,M_{\2 \2} +\, i \,\displaystyle{\frac{\epsilon^{\2} M_{\1 \2}
-\alpha^{\2}_{\mi}
 M_{\2 \1}}{\epsilon\,\alpha_{\mi}}}\,
 -  \left(M_{\1 \3} + i
M_{\2 \4} - \frac{\epsilon^{\2}M_{\1 \4} + i \,\alpha^{\2}_{\mi} M_{\2 \3}
}{\epsilon\,\alpha_{\mi}}\,\right)\, \frac{M_{\3 \1} - i M_{\4 \2} +
\displaystyle{ \frac{i\,\epsilon^{\2} M_{\3 \2} -\alpha^{\2}_{\mi}
 M_{\4 \1}}{\epsilon\,\alpha_{\mi}}}}{M_{\4 \4}+M_{\3 \3} - \displaystyle{
 \frac{\epsilon^{\2} M_{\3 \4} +
  \alpha^{\2}_{\mi}
 M_{\4 \3}}{\epsilon\,\alpha_{\mi}}}}\,\,.
\]
The explicit formulas of the elements of the matrix $M$  are
\begin{equation}
\label{melem}
\begin{array}{clcl}
(1-zw) & M_{\1 \1} & = &
+\cosh(\alpha_{\mi}\lambda) - \beta \gamma \cosh(\alpha_{\pl}\lambda) \,\,,\\
(1-zw) & M_{\1 \2} & = & - \sinh(\alpha_{\mi}\lambda) + \alpha_{\mi\pl}\, \beta \gamma \sinh(\alpha_{\pl}\lambda)\,\,,\\
(1-zw) & M_{\1 \3} & = & -\beta\,[\,\cosh(\alpha_{\mi}\lambda) - \cosh(\alpha_{\pl}\lambda)\,]\,\,, \\
(1-zw) & M_{\1 \4} & = & +\beta\,[\,\sinh(\alpha_{\mi}\lambda) - \alpha_{\mi\pl}\sinh(\alpha_{\pl}\lambda)\,]\,\,,\\
(1-zw) & M_{\2 \1} & = & - \sinh(\alpha_{\mi}\lambda) + \alpha_{\pl\mi}\, \beta \gamma \sinh(\alpha_{\pl}\lambda)\,\,,\\
(1-zw) & M_{\2 \2} & = & +\cosh(\alpha_{\mi}\lambda) - \beta \gamma \cosh(\alpha_{\pl}\lambda)\,\,, \\
(1-zw) & M_{\2 \3} & = & +\beta\,[\,\sinh(\alpha_{\mi}\lambda) - \alpha_{\pl\mi} \sinh(\alpha_{\pl}\lambda)\,]\,\,,\\
(1-zw) & M_{\2 \4} & = &-\beta\,[\,\cosh(\alpha_{\mi}\lambda) - \cosh(\alpha_{\pl}\lambda)\,]\,\,, \\
(1-zw) & M_{\3 \1} & = & +\gamma\,[\,\cosh(\alpha_{\mi}\lambda) - \cosh(\alpha_{\pl}\lambda)\,]\,\,,\\
(1-zw) & M_{\3 \2} & = & -\gamma\,[\,\sinh(\alpha_{\mi}\lambda) - \alpha_{\mi\pl}\,\sinh(\alpha_{\pl}\lambda)\,]\,\,, \\
(1-zw) & M_{\3 \3} & = & - \beta \gamma \cosh(\alpha_{\mi}\lambda) + \cosh(\alpha_{\pl}\lambda)\,\,,\\
(1-zw) & M_{\3 \4} & = &+\beta \gamma \sinh(\alpha_{\mi}\lambda) - \alpha_{\mi\pl}\,\sinh(\alpha_{\pl}\lambda)\,\,, \\
(1-zw) & M_{\4 \1} & = &-\gamma\,[\,\sinh(\alpha_{\mi}\lambda) - \alpha_{\pl\mi}\,\sinh(\alpha_{\pl}\lambda)\,]\,\,,  \\
(1-zw) & M_{\4 \2} & = &+\gamma\,[\,\cosh(\alpha_{\mi}\lambda) - \cosh(\alpha_{\pl}\lambda)\,] \,\,,\\
(1-zw) & M_{\4 \3} & = &+\beta \gamma \sinh(\alpha_{\mi}\lambda) - \alpha_{\pl\mi}\,\sinh(\alpha_{\pl}\lambda)\,\,,\\
(1-zw) & M_{\4 \4} & = &- \beta \gamma \cosh(\alpha_{\mi}\lambda)
+ \cosh(\alpha_{\pl}\lambda)\,\,,
\end{array}
\end{equation}
where $\alpha_{\pl\mi} = \alpha_{\pl}/\alpha_{\mi}$ and $\alpha_{\mi\pl}
=\alpha_{\mi}/\alpha_{\pl}$.

A first important result is  the evidence that the transmission
coefficient does {\em not} depend on $\theta$. This not trivial
result is due to the fact that the closed formula obtained for $T$
contains the terms $M_{\1\1,\1\2,\2\1,\2\2}$ and
$M_{\3\3,\3\4,\4\3,\4\4}$ which are independent of $\theta$, their
formulas contain the quantity $\beta \gamma$ which is independent
on $\theta$, and the terms proportional to $\beta$, i.e.
$M_{\1\3,\1\4,\2\3,\2\4}$, which are multiplied by terms
proportional to $\gamma$, i.e. $M_{\3\1,\3\2,\4\1,\4\2}$. The
invariance of the transmission coefficient in the plane
$(V_{\2}\,,\,V_{\3})$, seen in the numerical studies appeared in
litterature\cite{DDN02}, is now proved to be a property of
quaternionic quantum mechanics.

The results of complex quantum mechanics can be obtained  by
taking the limits $(V_c,V_q) \to (1,0)$. Observing that
$\alpha_{\pem}\to\sqrt{1\pm \epsilon^{\2}}$ and
 $(z,w) \to (0,0)$,  we  find
\begin{equation}
\label{complex}
\begin{array}{lcl}
M_{\1 \1} & = &
+\cosh(\sqrt{1- \epsilon^{\2}}\,\lambda)  \,\,,\\
M_{\1 \2} & = & - \sinh(\sqrt{1- \epsilon^{\2}}\,\lambda) \,\,,\\
 M_{\1 \3} & = & 0\,\,, \\
 M_{\1 \4} & = & 0\,\,,\\
 M_{\2 \1} & = & - \sinh(\sqrt{1- \epsilon^{\2}}\,\lambda) \,\,,\\
 M_{\2 \2} & = & + \cosh(\sqrt{1- \epsilon^{\2}}\,\lambda) \,\,, \\
 M_{\2 \3} & = & 0\,\,,\\
 M_{\2 \4} & = &0\,\,, \\
 M_{\3 \1} & = &0\,\,,\\
 M_{\3 \2} & = &0\,\,, \\
 M_{\3 \3} & = & + \cosh(\sqrt{1+ \epsilon^{\2}}\,\lambda)\,\,,\\
 M_{\3 \4} & = & - \sqrt{(1- \epsilon^{\2})/(1+ \epsilon^{\2})}\,\,
 \sinh(\sqrt{1+ \epsilon^{\2}}\,\lambda)\,\,, \\
 M_{\4 \1} & = &0\,\,,  \\
 M_{\4 \2} & = & 0 \,\,,\\
 M_{\4 \3} & = &  -\sqrt{(1+ \epsilon^{\2})/(1- \epsilon^{\2})}\,
 \,\sinh(\sqrt{1+ \epsilon^{\2}}\,\lambda)\,\,,\\
 M_{\4 \4} & = &
+ \cosh(\sqrt{1+ \epsilon^{\2}}\,\lambda)\,\,.
\end{array}
\end{equation}
In this limit, the expression for $\mathcal{D}$ reduces to
\[ \mathcal{D}_c= 2\,\cosh
(\sqrt{1-\epsilon^{\2}}\,\lambda) + i
\,\frac{1-2\,\epsilon^{\2}}{\epsilon
\sqrt{1-\epsilon^{\2}}}\,\sinh
(\sqrt{1-\epsilon^{\2}}\,\lambda)\,\,,
\]
and the transmission coefficient becomes\cite{COHEN,MER}
 \[
  T_c = e^{-\,i\epsilon \lambda} \,\mbox{\Large $/$}\,\left[\, \cosh
(\sqrt{1-\epsilon^{\2}}\,\lambda) + i
\,\frac{1-2\,\epsilon^{\2}}{2\,\epsilon
\sqrt{1-\epsilon^{\2}}}\,\sinh
(\sqrt{1-\epsilon^{\2}}\,\lambda)\,\right]\,\,.
\]
In the case of diffusion, $\epsilon>1$, the transmission
coefficient is often given in terms of cosine and sines functions,
i.e.
 \begin{equation}
 \label{trac2}
 T_c = e^{-\,i\epsilon \lambda} \,\mbox{\Large $/$}\,\left[\, \cos
(\sqrt{\epsilon^{\2}-1}\,\lambda) + i
\,\frac{1-2\,\epsilon^{\2}}{2\,\epsilon
\sqrt{\epsilon^{\2}-1}}\,\sin
(\sqrt{\epsilon^{\2}-1}\,\lambda)\,\right]\,\,.
\end{equation}

\section*{III. RESONANCE'S PHENOMENA}

In this section, we analyze when possible  analytically and when
not numerically the phenomenon of resonances. For the complex
case, the transmission probability
\begin{equation}
 \label{trapc}
 |T_c|^{^{2}} = \left[\,1 +
 \frac{\sin^{\2}(\sqrt{\epsilon^{\2}-1}\,\lambda)}{4\,\epsilon^{\2}(\epsilon^{\2}-1)}\,
 \right]^{^{-1}}\,\,,
\end{equation}
shows that the phenomenon of resonances, $|T_c|=1$, happens when the
energy/barrier width condition
\begin{equation}
\label{res} \sqrt{\epsilon_n^{\2}-1}\,\,\lambda_n = n\,\pi
\end{equation}
is satisfied\cite{COHEN,MER}. The transmission probability
oscillates between one and
\begin{equation}
 \label{trapc}
 |T_c|_{min}^{^{2}} = \left[\,1 +
 \frac{1}{4\,\widetilde{\epsilon}_n^{^{\,\,2}}(\widetilde{\epsilon}_n^{^{\,\,2}}-1)}\,
 \right]^{^{-1}}\,\,,
\end{equation}
obtained for
$\sqrt{\widetilde{\epsilon}_n^{^{\,\,2}}-1}\,\,\widetilde{\lambda}_n =
(2\,n+1)\,\pi\,/\,2$. This minimum tends to one for increasing incoming
energies.\\

\noindent $\bullet$ Fixing the width of the potential barrier,
$\lambda_n=\lambda_{\0}$, and varying the energy of the incoming particle, we
can find the energy values, $(\epsilon_n,\epsilon_n + \Delta \epsilon_n)$, for
which the transmission probability reaches two consecutive maxima, and the
energy value, $\epsilon_n + \Delta \widetilde{\epsilon}_n$ of the minimum
between such maxima,
\[
\begin{array}{rclcl}
\epsilon^{\2}_n -1 & = & n^{\2}\pi^{\2}\,/\,\lambda_{\0}^{{^2}}\,\,,\\
(\epsilon_n+\Delta \widetilde{\epsilon}_n)^{^{2}}-1 & = &
(n+1/2)^{^{2}}\pi^{\2}\,/\,\lambda_{\0}^{{^2}}&
\,\,\,\,\,\,\,\Rightarrow\,\,\,\,\,\,\, & \Delta \widetilde{\epsilon}_n=
\sqrt{1 + (n+1/2)^{^{2}}\,\pi^{\2}\,/\,\lambda_{\0}^{{^2}}} - \sqrt{1 +
n^{^{2}}\,\pi^{\2}\,/\,\lambda_{\0}^{{^2}}}\,\,, \\
(\epsilon_n+\Delta \epsilon_n)^{^{2}}-1 & = &
(n+1)^{^{2}}\pi^{\2}\,/\,\lambda_{\0}^{{^2}} &\Rightarrow  & \Delta
\epsilon_n= \sqrt{1 + (n+1)^{^{2}}\,\pi^{\2}\,/\,\lambda_{\0}^{{^2}}} -
\sqrt{1 + n^{^{2}}\,\pi^{\2}\,/\,\lambda_{\0}^{{^2}}}\,\, .
\end{array}
\]
For complex potential, $(V_c,V_q)=(1,0)$, and for a barrier width of $3\pi$,
we find
\[
\begin{array}{lcl}
\epsilon_{\1}=\sqrt{10}/\,3\sim 1.054&\,\,,\,\,\,\,\, &\Delta
\epsilon_{\1}=(\sqrt{13}-\sqrt{10})/\,3\sim 0.148\,\,,\\
\epsilon_{\2}=\sqrt{13}/\,3\sim 1.202 &\,\,,\,\,\,\,\, &\Delta
\epsilon_{\2}=\sqrt{2}-\sqrt{13}/\,3\sim 0.212\,\,,\\
\epsilon_{\3}=\sqrt{2}\sim 1.414\,\,.& &
\end{array}
\]

\noindent $\bullet$ Fixing the energy of the incoming particle,
$\epsilon_n=\epsilon_{\0}$, and varying the barrier width, we find
\[
\begin{array}{rclcl}
\lambda_n  & = & n\,\pi\,/\,\sqrt{\epsilon_{\0}^{{^2}}-1}\,\,,\\
 \lambda_n +\Delta \widetilde{\lambda} & = & (n+1/2)\,\,\pi\,/\,
 \sqrt{\epsilon_{\0}^{{^2}}-1}
& \,\,\,\,\,\,\,\Rightarrow\,\,\,\,\,\,\, & \Delta
\widetilde{\lambda}=\pi/2\,\sqrt{\epsilon_{\0}^{\2}-1}\,\,,
\\
\lambda_n +\Delta \lambda & = & (n+1)\,\,\pi\,/\,
 \sqrt{\epsilon_{\0}^{{^2}}-1}
& \,\,\,\,\,\,\,\Rightarrow\,\,\,\,\,\,\, & \Delta
\lambda=\pi/\sqrt{\epsilon_{\0}^{\2}-1}\,\,.
\end{array}
\]
For complex potential, $(V_c,V_q)=(1,0)$, and for an incoming
energy  $\epsilon_{\0}=\sqrt{2}$, we find
\[
\begin{array}{lcl}
\lambda_{\1}=2\,\pi&\,\,,\,\,\,\,\, &\Delta
\lambda_{\1}=\pi\,\,,\\
\lambda_{\2}=3\,\pi &\,\,,\,\,\,\,\, &\Delta
\lambda_{\2}=\pi\,\,,\\
\lambda_{\3}=4\,\pi\,\,.& &
\end{array}
\]
In Fig.\,1, where the barrier width is set to $3\pi$, we see the
phenomenon of resonance for complex, mixed and pure quaternionic
potentials varying the energy of the incoming particle. It is
interesting to observe that $\epsilon_n$ and $\Delta \epsilon_n$
decrease when the quaternionic part of the potential tends to one,
\[\begin{array}{|c||c|c|c|c|c|}\hline
(\,V_c\,,\,V_q\,) & \epsilon_{\1} & \epsilon_{\2} & \Delta
\epsilon_{\1} & \epsilon_{\3} & \Delta \epsilon_{\2}\\  \hline
\hline
 (\,1\,,\,0\,) & \,\,1.054\,\, & \,\,1.202\,\, & \,\,0.148\,\, &
 \,\,1.414\,\, & \,\,0.212\,\,\\ \hline \hline
(\, \sqrt{3}/2 \,,\,1/2\,) & 1.049 & 1.188 & 0.139 & 1.394 & 0.206\\
\hline  \hline (\,1/\sqrt{2}\,,\,1/\sqrt{2}\,) & 1.043 & 1.170 &
0.127 & 1.369 & 0.199 \\ \hline \hline (\,1/2\,,\,\sqrt{3}/2\,) &
1.034 & 1.145 & 0.111 & 1.334 &
0.189\\ \hline  \hline (\,0\,,\,1\,) & 1.011 & 1.077 & 0.066 & 1.246 & 0.169\\
\hline
\end{array}
\]
In Fig.\,2, where the incoming energy is set to $\sqrt{2}$, we see
the phenomenon of resonance for complex, mixed and pure
quaternionic potentials varying the barrier width. It is
interesting to observe that $\Delta \lambda$ is constant for a
fixed potential and  decreases when the quaternionic part of the
potential tends to one,
\[
\begin{array}{|c||c|c|c|c|c|}\hline
(\,V_c\,,\,V_q\,) & \lambda_{\1}[\pi] & \lambda_{\2}[\pi] & \Delta \lambda_{
\1}[\pi] & \lambda_{\3}[\pi] & \Delta \lambda_{\2}[\pi]\\  \hline \hline
 (\,1\,,\,0\,) & 2  & 3 & 1 & 4 & 1 \\ \hline \hline
(\, \sqrt{3}/2 \,,\,1/2\,) & \,\,1.949\,\, & \,\,2.915\,\, & \,\,0.966\,\, &
\,\,3.881\,\, & \,\,0.966\,\,\\
\hline  \hline (\,1/\sqrt{2}\,,\,1/\sqrt{2}\,) & 1.890 & 2.817 & 0.927 & 3.744
& 0.927 \\ \hline \hline (\,1/2\,,\,\sqrt{3}/2\,) & 1.819 & 2.695 & 0.876 &
3.571 &
0.876\\ \hline  \hline (\,0\,,\,1\,) & 1.718 & 2.478 & 0.760 & 3.238 & 0.760\\
\hline
\end{array}
\]

\section*{IV. THE LIMIT CASE BETWEEN DIFFUSION AND TUNNELING}

In this section, we present the interesting case of $\epsilon=1$, i.e. the
limit situation between diffusion and tunneling. We start by analyzing  this
limit for a complex potential $(V_c,V_q)=(1,0)$. In the free region, the
solution of the Schr\"odinger equation, for $\epsilon=1$, becomes
\begin{eqnarray*}
\Phi_{c,\I}^{\bullet}(\xi) & = & \exp \left[ \, i \, \xi \, \right] + \exp
\left[ \, - \, i \, \xi \, \right] \, R_c^{\bullet} + j \, \exp \left[
\, \xi \, \right] \, \widetilde{R}^{\bullet}_c\,\,, \nonumber \\
\Phi_{c,\III}^{\bullet}(\xi) & = & \exp \left[ \, i \, \xi \, \right] \,
T_c^{\bullet} + j \,
 \exp \left[ \, - \, \xi \,  \right] \, \widetilde{T}_c^{\bullet}\,\,.
\end{eqnarray*}
Eq.(\ref{qad}) reads
\begin{equation}
 i\,\partial_{\xi \xi}\,\Phi_{c,\II}^{\bullet}(\xi) = i\,\Phi_{c,\II}^{\bullet}(\xi)-
 \Phi_{c,\II}^{\bullet}(\xi)\,i\,\,,\,\,\,\,\,\,\,\,\,\,\,\,
0<\xi<\lambda\,\,.
\end{equation}
Separating the complex from the pure quaternionic part in the solution, i.e.
\[
\Phi_{c,\II}^{\bullet}(\xi)  =  \varphi_{c,\II}^{\bullet}(\xi)+ j\,
\psi^{\bullet}_{c,\II}(\xi)\,\,,
\]
we find two complex differential equations which lead to
\begin{eqnarray}
 \varphi_{c,\II}^{\bullet}(\xi) & = & A_c^{\bullet}\,\xi + B_c^{\bullet}\,\,, \nonumber \\
 \psi^{\bullet}_{c,\II}(\xi) & = &C_c^{\bullet}\,\exp[\,\sqrt{2}\,\xi\,] +
 D_c^{\bullet}\,\exp[\,-\,\sqrt{2}\,\xi\,]\,\,,
\end{eqnarray}
observe that  only two  coefficients appear in the  complex solution
$\varphi_{c,\II}^{\bullet}(\xi)$. Imposing the continuity conditions, we find
\begin{eqnarray}
R_{c}^{\bullet}&=&-\,i\,\lambda\,/\,(2-i\,\lambda)\,\,,\nonumber
\\
 T_{c}^{\bullet}&=&2\,\exp[\,-\,i\,\lambda\,]\,/\,(2-i\,\lambda)\,\,,
 \end{eqnarray}
and as expected $\widetilde{R}_{c}^{\bullet}=\widetilde{T}_{c}^{\bullet}=0$
reproducing the standard complex dynamics. In the case of pure quaternionic
potentials, the free solutions are
\begin{eqnarray*}
\Phi_{q,\I}^{\bullet}(\xi) & = & \exp \left[ \, i \, \xi \, \right] + \exp
\left[ \, - \, i \, \xi \, \right] \, R_q^{\bullet} + j \, \exp \left[
\, \xi \, \right] \, \widetilde{R}^{\bullet}_q\,\,, \nonumber \\
\Phi_{q,\III}^{\bullet}(\xi) & = & \exp \left[ \, i \, \xi \, \right] \,
T_q^{\bullet} + j \,
 \exp \left[ \, - \, \xi \,  \right] \, \widetilde{T}_q^{\bullet}\,\,,
\end{eqnarray*}
and the equation to be solved is
\begin{equation}
\label{sss}
 i\,\partial_{\xi \xi}\,\Phi_{q,\II}^{\bullet}(\xi) = j
\,e^{-i\theta}\,\Phi_{q,\II}^{\bullet}(\xi)-
 \Phi_{q,\II}^{\bullet}(\xi)\,i\,\,,\,\,\,\,\,\,\,\,\,\,\,\,
0<\xi<\lambda\,\,.
\end{equation}
Separating the complex from the pure quaternionic part,
\begin{equation} \Phi_{q,\II}^{\bullet}(\xi)  =
\varphi_{q,\II}^{\bullet}(\xi)+ j\, \psi^{\bullet}_{q,\II}(\xi)\,\,,
\end{equation}
we find, as solutions of the two complex differential equations coming from
Eq.(\ref{sss}),
\begin{eqnarray}
\varphi^{\bullet}_{q,\II}(\xi)  &=& A_q^{\bullet}\,\xi^{^{3}} +
B_q^{\bullet}\,\xi^{\2} + C_q^{\bullet}\,\xi + D_q^{\bullet}\,\,,\nonumber \\
\psi_{q,\II}(\xi)  &=& -\,i\,e^{-i\theta}\left[\,A_q^{\bullet}\,\xi^{^{3}} +
B_q^{\bullet}\,\xi^{\2}+ (6\,A_q^{\bullet}+C_q^{\bullet})\, \xi +
2\,B_q^{\bullet}+D_q^{\bullet}\right]\,\,.
\end{eqnarray}
By imposing the continuity conditions, we obtain the reflection and
transmission amplitudes for pure quaternionic potentials in the limit case
between diffusion and tunneling,
\begin{eqnarray}
R_q^{\bullet} & = & -\,i\,\lambda^{^{2}}\,(6+4\,\lambda+\lambda^{^{2}})\,/\,
[\,24+24\,(1-i)\,\lambda-18\,i\, \lambda^{^{2}} -
4\,(1+i)\,\lambda^{^{3}}-\lambda^{^{4}}\,]\,\,,
\nonumber\\
T_q^{\bullet} &  =  & 2\,\exp[-i\lambda]\,
(12+12\,\lambda+6\,\lambda^{^{2}}+\lambda^{^{3}})\,/\,[\,24+24\,(1-i)\,\lambda-18\,i\,
\lambda^{^{2}} - 4\,(1+i)\,\lambda^{^{3}}-\lambda^{^{4}}\,]\,\,.
\end{eqnarray}
The expression obtained for $T_q^{\bullet}$ is in perfect
agreement with the limit for $\epsilon \to 1$ of the coefficient
$T$ given in Eq.(\ref{tram}).

The qualitative differences between complex and pure quaternionic potentials
can be, for example, seen comparing the reflection and transmission
coefficients in the case of thin ($\lambda\ll 1$) and large ($\lambda\gg 1$)
barriers. The computation taking into account fourth order corrections gives
\[
\begin{array}{clcl}
\lambda\ll 1\,\,:\,\,\,
 & |R_c^{\bullet}| & \,\,\,\sim\,\,\, & \lambda/2-\lambda^{^{3}}/16+
 \mbox{O}[\,\lambda^{^{5}}\,]\,\,,\\
 & |T_c^{\bullet}| & \,\,\,\sim\,\,\, & 1-\lambda^{^{2}}/8+3\lambda^{^{4}}/128+
 \mbox{O}[\,\lambda^{^{5}}\,]\,\,,\\ \\
 & |R_q^{\bullet}| & \,\,\,\sim\,\,\, & \lambda^{^{2}}/4-\lambda^{^{3}}/12+
 \mbox{O}[\,\lambda^{^{5}}\,]\,\,,\\
 & |T_q^{\bullet}| & \,\,\,\sim\,\,\, & 1-\lambda^{^{4}}/32+
 \mbox{O}[\,\lambda^{^{5}}\,]\,\,,\\ \\
\lambda\gg 1\,\,:\,\,\,
 & |R_c^{\bullet}| & \,\,\,\sim\,\,\, & 1 - 2/\lambda^{^{2}}+6/\lambda^{^{4}}+
 \mbox{O}[\,1/\lambda^{^{5}}\,]\,\,,\\
 & |T_c^{\bullet}| & \,\,\,\sim\,\,\, & 2/\lambda-4/\lambda^{^{3}}+
 \mbox{O}[\,1/\lambda^{^{5}}\,]\,\,,\\ \\
 & |R_q^{\bullet}| & \,\,\,\sim\,\,\, & 1-2/\lambda^{^{2}}-8/\lambda^{^{3}} +
 6/\lambda^{^{4}}+\mbox{O}[\,1/\lambda^{^{5}}\,]\,\,,\\
 & |T_q^{\bullet}| & \,\,\,\sim\,\,\, & 2/\lambda+4/\lambda^{^{2}}-8/\lambda^{^{3}}
 -8/\lambda^{^{4}}+
 \mbox{O}[\,1/\lambda^{^{5}}\,]\,\,.
\end{array}
\]

\section*{V. CONCLUSIONS}

With the development of a consistent theory of quaternionic differential
equations\cite{DeSco00,DeDuc01,DeScoSol02,DeDuc03,DeDuc04,DeDuc06}, it is now
possible to be more profound in discussing quaternionic quantum mechanics. The
problem of diffusion by a quaternionic potential, limited up to now to
numerical analysis\cite{PER79,KAI84,KLE88,DAV89,DAV92,DDN02,DD05,DD06,DD07},
was solved by a matrix approach and leaded to a closed formula for the
transmission amplitude.

As in the case of standard quantum mechanics, for incoming particle of energy
$\epsilon>1$, quaternionic barriers present the phenomenon of resonances. A
comparison between the complex, $V_c=1$, and the pure quaternionic, $V_q=1$,
cases show that, for pure quaternionic potentials,  the minimum value of
transmission increases whereas the oscillation interval decreases.

 This paper could motivate the study of quaternionic quantum mechanics by
 using the wave packet formalism. An interesting application is the study
 of tunneling times\cite{TT1,TT2}.  The Hartman effect\cite{HE} surely
  represents one of the most
 intriguing challenges recently appeared in literature\cite{TT3}. Quaternionic deviations
 from the standard phase, which is fundamental in the calculation of  tunneling times,
  can be now investigated by using the phase of the transmission coefficient
 given in this paper.

The analogy between quantum mechanics and the propagation of light
through stratified media\cite{DR08} suggests to analyze in details
the limit case $\epsilon=1$. In this case the momentum
distributions should be centered in $E_{\0}=V_{\0}$ and phenomena
of diffusion and tunneling must be treated together. The outgoing
wave packets should move with different velocity and should give
life to new interesting phenomena of interference.\\

\noindent{\bf Acknowledgements}. One of the authors (SdL) wish to
thank the Department of Physics, University of Salento (Lecce,
Italy), where the revised version of the paper was prepared, for
the invitation and the hospitality. SdL also thanks the FAPESP
(Brazil) for financial support by the grant n. 10/02216-2.

\newpage

\begin{figure}[hbp]
\hspace*{-2.5cm}
\includegraphics[width=19cm, height=24cm, angle=0]{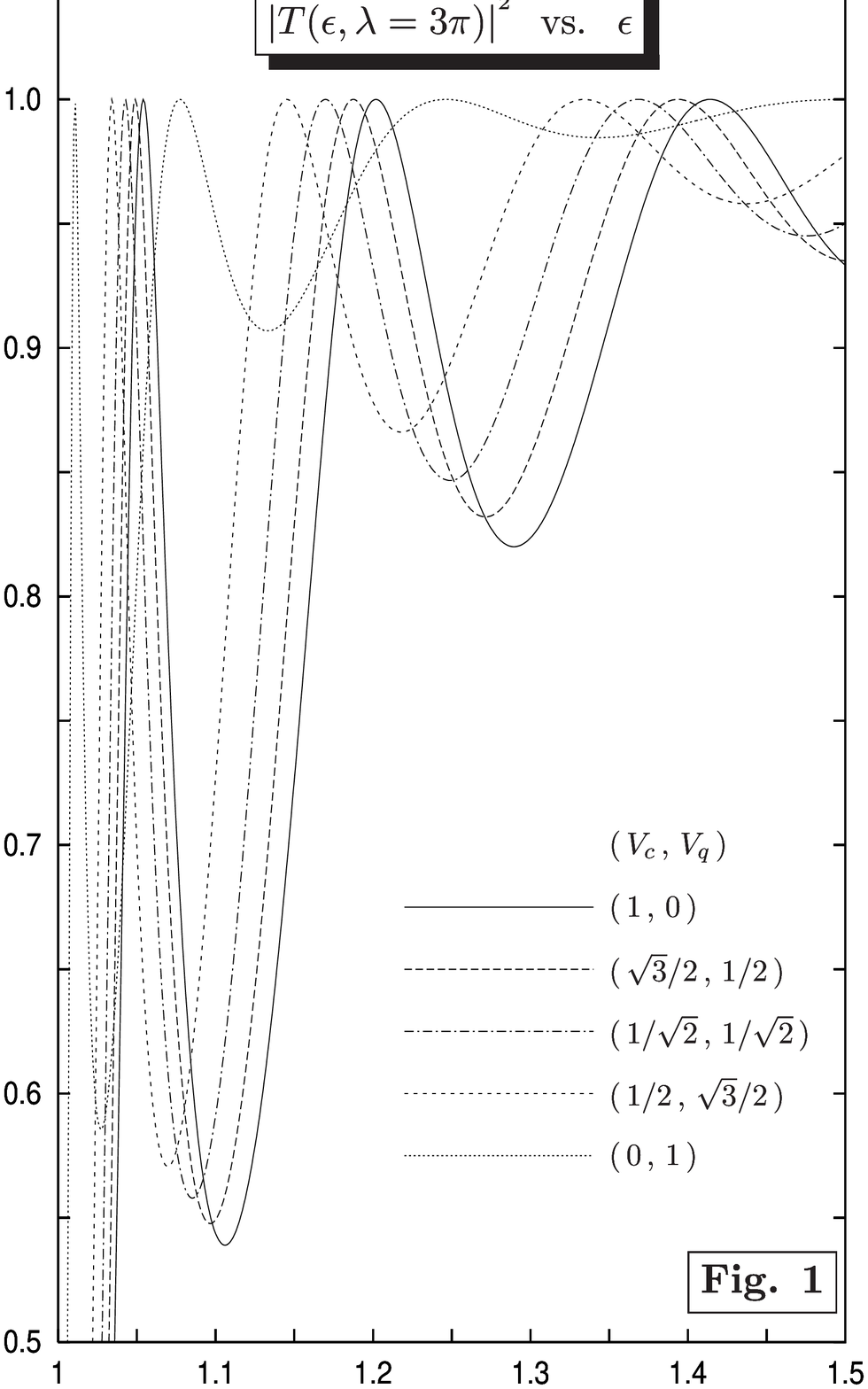}
\vspace*{-1.8cm}
 \caption{Incoming energy dependence of the transmission probability for different
 values of the potential. The continuous line represents the standard case of a
 complex potential. The dotted line represents the case of a pure quaternionic
 potential. The intermediate cases, show larger values for the minima of the
 transmission probability  when the quaternionic part of the potential increases.}
\end{figure}

\newpage

\begin{figure}[hbp]
\hspace*{-2.5cm}
\includegraphics[width=19cm, height=24cm, angle=0]{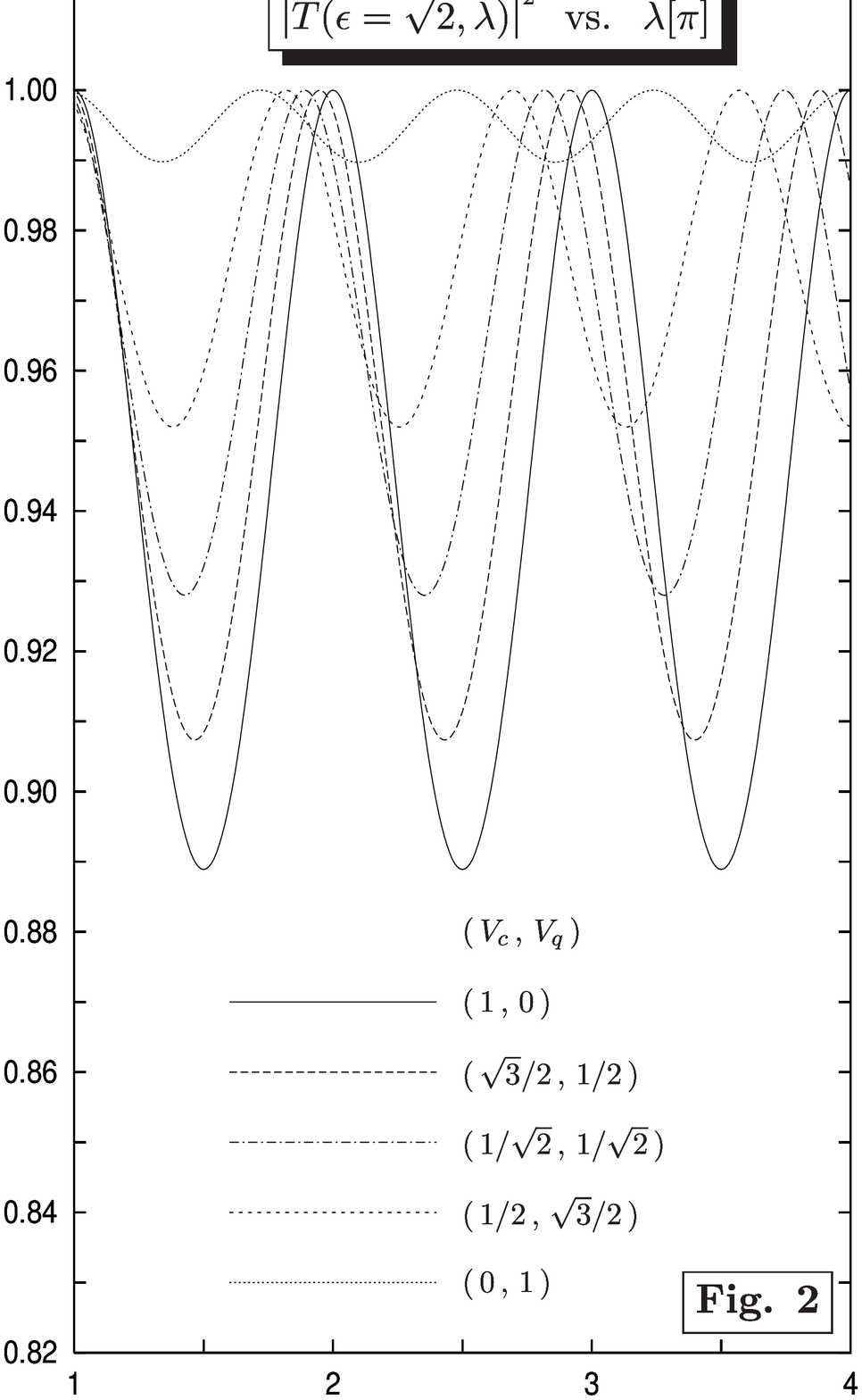}
\vspace*{-1.8cm}
 \caption{Barrier width dependence of the transmission probability for different
 values of the potential. The continuous line represents the standard case of a
 complex potential. The dotted line represents the case of a pure quaternionic
 potential. The intermediate cases, show a shorter oscillation period when the
 quaternionic part of the potential increases.}
\end{figure}

\end{document}